\documentclass[journal,]{IEEEtran}

\usepackage{graphicx,amsmath,amssymb,amsfonts,xcolor,soul,circuitikz,subfigure,placeins, multirow,stfloats, booktabs, comment}%
\usepackage[colorlinks,allcolors=black,citecolor=blue, urlcolor=blue]{hyperref}

\usepackage{url}
\usepackage{float}
\floatstyle{ruled}
\newfloat{model}{thp}{lop}
\floatname{model}{\textsc{Model}}

\DeclareMathOperator*{\minimize}{minimize}

\def\ModelOne{\mbox{\small{\textsc{\textsf{Model \ref{mod:NLPModel}}}}}}
\def\ModelTwo{\mbox{\small{\textsc{\textsf{Model \ref{Mod: CCModel}}}}}}
\def\ModelED{\mbox{\small{\textsc{\textsf{Model \ref{Mod: EDModel}}}}}}

\def\CaseA{\mbox{\small{\textsc{\textsf{Case NLP}}}}}
\def\CaseB{\mbox{\small{\textsc{\textsf{Case LP--Ideal}}}}}
\def\CaseC{\mbox{\small{\textsc{\textsf{Case LP--Approx}}}}}
\def\CaseD{\mbox{\small{\textsc{\textsf{Case MILP}}}}}

\definecolor{G}{RGB}{174,1,126}
\definecolor{new}{RGB}{35,132,67}
\setstcolor{red}

\ifCLASSINFOpdf
\else
\fi
\usepackage{amsmath}
\usepackage{array}
\begin{document}
%
\title{Non-ideal Linear Operation Model for Li-ion Batteries}
%
%
%

\author{\IEEEauthorblockN{Alvaro Gonzalez-Castellanos~,\IEEEmembership{Student Member,~IEEE},
David Pozo~,\IEEEmembership{Senior Member,~IEEE},
Aldo Bischi}}

%
%

\markboth{Journal of \LaTeX\ Class Files,~Vol.~14, No.~8, August~2015}%
{Shell \MakeLowercase{\textit{et al.}}: Bare Demo of IEEEtran.cls for IEEE Journals}
%



\maketitle
\thispagestyle{plain}
\pagestyle{plain}

\begin{abstract}
Currently, the characterization of electric energy storage units used for power system operation and planning models relies on two major assumptions: charge and discharge efficiencies, and power limits are constant and independent of the electric energy storage state of charge. This approach can misestimate the available storage flexibility.

This work proposes a detailed model for the characterization of steady-state operation of Li-ion batteries in optimization problems.
The model characterizes the battery performance, including non-linear charge and discharge power limits and efficiencies, as a function of the state of charge and requested power. We then derive a linear reformulation of the model without introducing binary variables, which achieves high computational efficiency, while providing high approximation accuracy. The proposed model characterizes more accurately the performance and technical operational limits associated with Li-ion batteries than those present in classical ideal models.

The developed battery model has been compared with three modelling approaches: the complete non-convex formulation; an ideal model typically used in the power system community; and a mixed integer linear reformulation approach. The models have been tested on a network-constrained economic dispatch for a 24-bus system. Based on the simulations, we observed approximately 12\% of energy mismatches between schedules that use an ideal model and those that use the model proposed in this study.
\end{abstract}


%
\IEEEpeerreviewmaketitle

\section*{Nomenclature}
\addcontentsline{toc}{section}{Nomenclature}
\begin{IEEEdescription}[\IEEEusemathlabelsep\IEEEsetlabelwidth{$R_{dif,elec/mem}$}]
\item[\textit{Indexes}]
\item[$g$] Generation unit.
\item[$l$] Power line.
\item[$n$] Power node.
\item[$t$] Time step.
\item[$j,k$] Indexes for characterization sample sets $J$ and $K$.
\item
\item[\textit{Parameters}]
\item[$A_{k}$] Interaction parameters for Redlich-Kister equation [J$\cdot$ mol$^{-1}$].
\item[$A_{nl}$] Line-to-node incidence matrix.
\item[${A}_\text{SEI}$] Area of solid-electrolyte interface [$\text{m}^2$].
\item[$C$] Cost [\$/Wh].
\item[$\chi_{and/ctd}$] Anode/catode molar fraction.
\item[$\Delta$] Size of time step [h].
\item[$\underline{\delta}/\overline{\delta}$] Minimum/Maximum angle allowed [rad].
\item[${E}_\text{A}$] Activation energy [$\text{kJ}\cdot \text{mol}^{-1}$].
\item[$\eta_{c}$] Coulombic efficiency.
\item[$\eta^\text{cha/dis}$] Charge/discharge efficiency.
\item[$\overline{E}$] Battery energy capacity [Wh].
\item[$\overline{F}$] Maximum power flow [W].
\item[$F$] Faraday constant [$\text{s}\cdot \text{A} \cdot \text{mol}^{-1}$].
\item[$\Gamma_{s,n}$] Battery-to-node incidence matrix.
\item[$i$] Current [A].
\item[${k}_0$] Reaction rate constant [$\text{m} \cdot \text{s}^{-1}$].
\item[$\Omega_{g,n}$] Generator-to-node incidence matrix.
\item[$P$] Power [W].
\item[$\underline{P}/\overline{P}$] Minimum/maximum power [W].
\item[$R$] Gas constant, [$\text{J}\cdot \text{mol}^{-1}\text{K}^{-1}$].
\item[$R_{ct}$] Charge transfer equivalent resistance [$\Omega$].
\item[$R_{dif,elec/mem}$] Electrode/Membrane diffusion equivalent resistance [$\Omega$].
\item[$R_{ohm}$] Ohmic losses equivalent resistance [$\Omega$].
\item[$SOC$] State of charge [p.u.].
\item[$SOC_{sur}$] State of charge at electrodes' surface [p.u.].
\item[$T$] Temperature [K].
\item[$U_{bat,0}$] Reference equilibrium potential [V].
\item[$v_{eq}$] Equilibrium voltage [V].
\item[$v_{INT}$] Non-ideal interaction voltage [V].
\item[$v_{INT,and/ctd}$] Non-ideal anode/catode interaction voltage [V].
\item[$X_{l}$] Series reactance in the line $l$ [p.u.].
\end{IEEEdescription}

%
\IEEEpeerreviewmaketitle

\section{Introduction}
\IEEEPARstart{T}{he} need for a secure and flexible operation of electric power systems and the falling prices of batteries has made large-scale Electric Energy Storage (EES) systems a viable and widely studied option. In particular, Li-ion battery systems have gained considerable attention because of their high energy density, power ratings, efficiency, and long lifetime \cite{Schimpe2018}. As an example, in South Italy and South Australia,  40 MW and 100 MW storage systems have been installed, respectively \cite{TernaReport,Lund2016a}. The introduction of EES, alongside renewable energy generation, compels the development of tools that can manage these systems optimally for maximum grid reliability and profitability \cite{Fares2014}.

The two major applications of EES that help achieve higher profit potential include frequency regulation and load shifting \cite{Gunter2016}. Multiple market mechanisms have been proposed for providing services to the power grid from EES: bulk energy storage, vehicle-to-grid (V2G), and distributed battery networks. V2G enables the use of electric vehicles to provide the aforementioned services, in addition to demand shifting via the smart charging of vehicles \cite{Uddin2018}. Domestic batteries can also be used for grid-scale services through their aggregation \cite{Rappaport2017}. Given the fast response needed for primary frequency control applications (in the range of seconds), battery systems would require the modeling of their dynamic processes. On the other hand, load shifting applications assume steady state behavior, i.e., constant parameters during the operational time steps considered, from minutes to hours.

Battery models can be divided into mathematical, electrochemical, and equivalent-circuit \cite{Rodriguez2017} models. The first category, \textit{mathematical models}, describes battery behavior based on the state of charge - $SOC$ (i.e., the ratio between stored energy and battery capacity), state of health (the battery's ability to perform in comparison to manufacturer specifications), and other macroscopic properties. Computational methods are used to derive mathematical models with a small number of variables, resulting in low computational costs, but without reflecting the internal processes in the cells, e.g., the changes on the equivalent voltage as a function of the stored energy. The second category, \textit{electrochemical models}, describes the chemical behavior of battery cells based on the physical and chemical processes that occur in the battery cells. Because of their accuracy, electrochemical models can be applied to the optimization of processes that are related to the design of the cell's physical parameters. Additionally, reduced-order electrochemical models allow the accurate modelling of the electrochemical processes in the battery, while providing a computationally efficient storage characterization for online control applications \cite{Rodriguez2017}. However, dynamic models are not suitable for their use in optimization applications which require the analysis of large time horizons. The third category, \textit{equivalent-circuit models}, provides an equivalent representation of the battery cell based on laboratory measurements. The electric elements in the circuit can be denoted by a combination of linear and exponential models to attune for the dynamic processes in the battery. The resultant circuits provide computationally efficient models, without requiring time-consuming laboratory measurements for the estimation of battery parameters \cite{Ali2017}.

\textit{Vagropoulos et al.} developed a linear optimization model for the optimal operation of an electric vehicle aggregator \cite{Vagropoulos2013}. The model is based on battery charging processes at constant efficiency, which are divided into two stages: constant current and constant voltage charge. 
\textit{Pand\v{z}i\'c} et al., provided piecewise linearized charging limits based on the battery's stored energy levels \cite{Pandzic2019}. Their model is based on a constant-current-constant-voltage charging cycle and validated based on laboratory experiments.
The effect of the requested current in storage efficiency is modeled by \textit{Wang et al.} \cite{Wang2016}. A concave and monotonically increasing function is provided to represent battery efficiency in terms of deviations from the system's reference current.

A detailed characterization of the loss mechanisms prevalent in both battery and power electronics is provided by \textit{Schimpe et al.} \cite{Schimpe2018}.
An analysis of the efficiency changes as a function of the stored energy level and power request is provided by \textit{Morstyn et al.} \cite{Morstyn2018}. The energy storage system characterization is approximated as a second-order model for its application in model predictive control for distributed microgrids with photovoltaic generation.

\textit{Ali et al.} developed a methodology for the non-linear estimation of Li-ion battery parameters in an equivalent circuit model \cite{Ali2017}. The estimated parameters reflect the dynamic processes by characterizing the elements in the equivalent circuit model as a function of the $SOC$.
\textit{Berrueta et al.} developed an equivalent circuit model for a Li-ion battery based on experimental data and the underlying electrochemical phenomena that characterize its performance \cite{Berrueta2018}. The proposed model achieves high accuracy while being computationally simple to implement. Additionally, the model characterizes the battery by the use of its state variables: cell temperature, current, and SOC. An efficiency-based equivalent-circuit model has been proposed by \textit{Rampazzo et al.} \cite{Rampazzo2018}. The model simulates the battery performance based on the battery state during the operation, as in \cite{Berrueta2018}. A Mixed Integer Linear Programming (MILP) model for representing the behavior of a Li-ion battery pack based on the battery's electrochemical behavior was developed by \textit{Sakti et al.} \cite{Sakti2017}. In this MILP model, the power limits and battery efficiency are expressed as a function of the $SOC$ and the power output. The nonlinearities present in the characterization are addressed through piecewise linearization based on simulated sample points.\\
%
%
\subsection{Paper Contributions and Organization}
The primary contribution of this study is the development of a detailed linear Li-ion battery model for its use in the operation optimization of power systems. This work takes into account the use of energy storage for economic operations in which the time steps considered are in the order of minutes/hours. Therefore the transient characterization of the batteries is assumed negligible. The proposed methodology models the operation limits and efficiencies of Li-ion batteries based on the characteristic charging and discharging curves. These characteristic curves can be obtained based on either computational simulations or by direct measurements, which allows the proposed model to be adapted to different energy storage technologies and updated when new simulated/measured data is available.\\
For this purpose we devote  Section II for introducing an equivalent circuit representation for the electrochemical behavior of Li-ion batteries, inspired by the work of \textit{Berrueta et al.} \cite{Berrueta2018}.  In Section III we  introduce a detailed non-linear model extended from the described equivalent circuit for the state-dependent charge, and discharge processes of Li-ion batteries based on the characterization of battery efficiencies and power limits. Thirdly, in Section IV, we propose a linear approach for battery operation through the convex curves definition of detailed non-linear battery models. Lastly, Section V and VI integrate our proposed Li-ion battery model into a day-ahead economic dispatch to illustrate the benefits of a detailed linear battery model compared with: the non-convex formulation of Section II, existing ideal model approaches (see, e.g., \cite{Pozo2014}), and approximation of non-convex curves through piecewise linearization based on mixed-integer programming. The computational and reliability benefits of implementing the developed model are also quantified.
\section{Battery Electrochemical Model}\label{Sect: Battery Physics}
A Li-ion battery reversibly stores electric energy as electrochemical energy. The positive electrode, cathode, of a Li-ion battery is composed of metal oxide materials, usually transition metals, and the negative electrode, anode, is made of graphite. When the battery is being charged, the Li ions flow from the cathode to the anode, where they combine with the incoming electrons and are later stored in the graphite layers \cite{Astaneh2018}.
During the discharge process, electrons are transferred from the battery cell, and the Li ions flow through the electrolyte back to the cathode. 

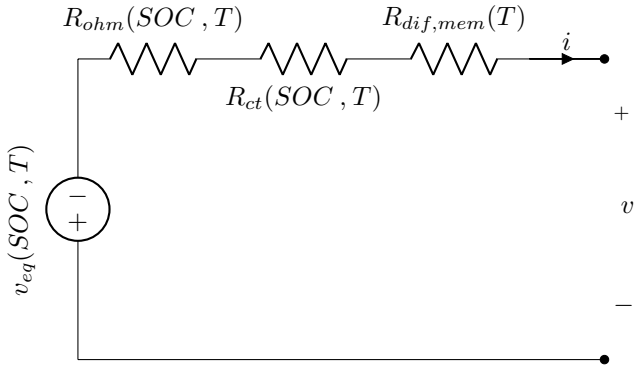
\begin{figure}[h] 
\begin{center}
\begin{circuitikz}[scale = 1, american voltages]
\draw (-0.75,3) node[rotate=90,left, color = black]{ $v_{eq}(SOC\mathbin{,} T)$};
\draw
	(0,4) to [american voltage source, v_<] (0,0)
    (0,4) to [R, l^={$R_{ohm}(SOC\mathbin{,} T)$}] (2,4)
    to [R, l_={$R_{ct}(SOC\mathbin{,} T)$}] (4,4)
  	to [R,l^={$R_{dif,mem}(T)$}] (6,4)
    to [short, i^>=$i$] (7,4)
  	to [open, v^>={$v_{}$}, *-*] (7,0)
  	to [short] (0,0)
 	 ; 
  \end{circuitikz}
\end{center}
\caption{Steady-state battery equivalent circuit}
\label{Fig: Circuit}
\end{figure}

After its dynamic processes have stabilized, a Li-ion battery can be represented as an equivalent resistive circuit (Fig. \ref{Fig: Circuit}). The equivalent-circuit representation consists of a voltage source associated with the electrochemical equilibrium voltage and three resistors that represent different electrochemical processes: ohmic losses, charge transfer, and membrane diffusion. In this section we describe each of the elements and its correspondence with the underlying electrochemical phenomena based on the experimentally validated model of \textit{Berrueta et al.} \cite{Berrueta2018}.

\subsection{Equilibrium voltage}
The equilibrium voltage, $v_{eq}$, indicates the difference in electrochemical potential (voltage) between the electrodes after the charge transfer dynamic processes have reached a steady state. Battery equilibrium voltages can be expressed as a summation of the three processes inside the cell, \eqref{Eq: v_eq}. The cell reference potential is defined by $U_{bat,0}$, which indicates the cell potential at standard concentrations. The concentration, i.e., molar fraction $\chi$, of the reactants on the electrodes changes with the amount of stored energy, which is captured by expressions \eqref{Eq: chi_a} and \eqref{Eq: chi_c}. Therefore, there is a change in the cell potential for nonstandard concentrations obtained by the second term in the summation. The combination of the first two terms gives the Nernst equation for nonstandard conditions (reactions not occurring at $298.15 K$, $1$ atmosphere, or a cathode and anode molarity of $1.0 M$). The third term, $v_{INT}$, reflects the non-ideal interactions between Li ions and the host matrix. The non-ideal interactions can be calculated based on a Redlich-Kister polynomial equation of seventh order, \eqref{Eq: Redlich-Kister}.
\begin{IEEEeqnarray}{rClr}
{v_\text{eq}(SOC{,}T)} &{{=}}& {U_\text{bat,0} {+} \frac{RT}{F} \cdot ln\left( \frac{ (1- \chi_{\text{ctd}} ) \cdot \chi_{\text{and}}}{\chi_{\text{ctd}} {\cdot} (1- \chi_{\text{and}} )} \right) {+} v_\text{INT}} \label{Eq: v_eq} \IEEEeqnarraynumspace \\
\noalign{\noindent where
\vspace{2\jot}}
&&\chi_{\text{and}} = 0.083 + 0.917 \cdot SOC,\IEEEyessubnumber* \label{Eq: chi_a}\\
&&\chi_{\text{ctd}} = 1 - 0.7 \cdot SOC,\label{Eq: chi_c}\\
&&v_\text{INT} = v_\text{INT,\text{ctd}} - v_\text{INT,\text{and}} \label{Eq: v_INT},
\end{IEEEeqnarray}
and
\begin{IEEEeqnarray}{rCll}
v_{\text{INT},j} &{=}& \sum\limits_{k=1}^7{A_k} \Bigg[ (2\chi_{j}{-}1)^{k} {-} \frac{2 \chi_{j}(k{-}1)(1-\chi_{j})}{(2\chi_{j}{-}1)^{2-k}} \Bigg]{,}
\IEEEyessubnumber* \label{Eq: Redlich-Kister}\\
&&&j{=} \;\text{and},\text{ctd}. \IEEEnonumber
\end{IEEEeqnarray}
\color{black}
\vspace{-3\jot}\subsection{Resistive elements}
The resistive elements $R_\text{ohm}$ and $R_{ct}$ indicate fast-dynamic processes. The former is associated with the \textit{ohmic phenomena} that represents the losses related to the movement of electrons and ions during charging and discharging processes; it depends linearly on the SOC and temperature, \eqref{eq:R_ohm}. The latter, $R_{ct}$, models \textit{charge transfer} through the solid-electrolyte interface (SEI), \eqref{eq:R_ct}. The SEI serves as a barrier between the electrodes and the electrolyte solution, preventing their spontaneous reaction (short-circuit) and enabling battery charge reversibility.
\begin{IEEEeqnarray}{l}
{R_\text{ohm}(SOC{,}T) = R_\text{ohm,0} + R_\text{ohm,T}{\cdot} T + R_\text{ohm,SOC} \cdot SOC   \label{eq:R_ohm}} \\
{R_{ct}(SOC{,}T) {=} \frac{1}{(\chi_{\alpha,and} {\cdot} \chi_{\alpha,ctd})^{0.5}} \cdot {\Bigg[} \frac{R \cdot T \cdot e^{E_A/R \cdot T}}{F^2\cdot A_{SEI} \cdot k_0} {\Bigg]} \label{eq:R_ct}}
\end{IEEEeqnarray}

The last resistive element of the equivalent-circuit model is related to the diffusion of Li ions through the membrane. The diffusion process causes a voltage drop in the battery cell and is inversely proportional to the operating temperature  \eqref{eq:R_diff,mem}. Similar to the membrane diffusion process, there exists an \textit{electrode diffusion} mechanism \eqref{eq:R_diff,elec}.
\begin{IEEEeqnarray}{l}
{R_{dif,mem}(T) = K_{dif,mem} \cdot \exp \Bigg( \frac{b_{dif,mem}}{T-T_{0,dif,mem}} \Bigg) \label{eq:R_diff,mem} \IEEEeqnarraynumspace}\\
{R_{dif,elec}(T) = K_{dif,elec} \cdot \exp\Bigg( \frac{b_{dif,elec}}{T-T_{0,dif,elec}}  \Bigg) \label{eq:R_diff,elec}}
\end{IEEEeqnarray}
Changes in the concentration of lithium on the electrodes results in a new perceived state of charge at the cell's electrode surface, $SOC_{sur}$. The difference between the $SOC$ and $SOC_{sur}$ results in a voltage drop that can be obtained in terms of the electrode diffusion process, \eqref{Eq: SOC_sur}. 
\begin{IEEEeqnarray}{rl}
SOC_{sur} & = SOC - R_{dif,elec} \cdot i \cdot \eta_c \label{Eq: SOC_sur} \\
\noalign{\noindent where the coulombic efficiency, $\eta_c$, is given by
\vspace{2\jot}}
\eta_c   & = \eta_{c,0} + \eta_{c,T} \cdot T + \eta_{c,i} \cdot i \label{Eq: Coulombic Eff}
 \end{IEEEeqnarray}
\label{App: PhysBatt}

In summary, all of the components present in the equivalent steady-state circuit model depend either on the temperature, $T$, the state of charge, $SOC$, the current, $i$, or a combination of all factors.

For the calculations presented in the following sections, the battery modeled will be that presented in \cite{Berrueta2018}, with 40 Ah and 133 V. A constant battery temperature of 25$^{\circ}$C will be assumed. The $SOC$ and the current will be considered control variables because they directly relate to the stored energy and the power requested from the battery. The values and the characterization of the battery parameters presented in this section, and used in the rest of the work, can be found in \cite{Berrueta2018}.
\section{Mathematical Battery Characterization} \label{Section: Techno Economic}
In this section, we present an alternative formulation to the classical ideal battery model used in the operation optimization of power systems (see e.g., \cite{Pozo2014}) by incorporating features that represent the internal electrochemical processes at the battery cell level. Consequently, we provide a more accurate description of battery behavior that can be used for power system economic operation. In doing so, charging and discharging power and SOC limits must be derived. However, instead of providing independent limits, as in \cite{Pozo2014}, we derive the limits from internal battery processes. The resulting mathematical battery model is of higher-dimension (more variables are needed), and it is non-linear and non-convex (a non-desired property for power system optimization models). On the other hand, it provides a more accurate mathematical description of the battery that is useful for power system optimization models. In this section, we gradually introduce the model, while in the next section, we propose a convex (linear) approach for the battery model.

\subsection{Power Limits \label{Sec: Limits}}
The $SOC$ at the electrode surface, $SOC_{sur}$,  is modeled based on \eqref{Eq: SOC_sur} and \eqref{Eq: Coulombic Eff} as:
\begin{IEEEeqnarray}{C}
SOC_{sur} = SOC {-} R_{dif,elec} {\cdot} i {\cdot}[\eta_{c,0} {+} \eta_{c,T} \cdot T {+} \eta_{c,i} {\cdot} i],
\IEEEeqnarraynumspace
\label{Eq: SOC_sur Limits}\\
\noalign{\noindent where
\vspace{2\jot}}
0\leq SOC_{sur} \leq 1. \label{Eq: SOC_sur bounded}
\end{IEEEeqnarray}
During the discharging process, the state of charge at the electrodes' surface $SOC_{sur}$ decreases; whereas in the charging one, it increases. Therefore the $SOC_{sur}$ lower bound is active during the discharge (but not its upper bound). Similarly,  the $SOC_{sur}$ upper bound is active during the charge. The calculation of the discharging limits is then derived as follows
\begin{IEEEeqnarray}{C}
0 \leq SOC_{sur}^\text{cha} \label{Eq: SOC_cha leq 1},\\
\noalign{\noindent where the maximum discharging current $\overline{I}^\text{dis}_0$ is the solution of the equation
\vspace{2\jot}}
SOC_{sur}^\text{cha}\left( \overline{I}^\text{dis}_0\right) = 0. \label{Eq: Imax_SOCsur}
\end{IEEEeqnarray}
We can obtain an explicit solution, in closed form, derived for the maximum discharging and charging current, $\overline{I}^\text{dis}_0$ and $\overline{I}^\text{cha}_0$, from the second-order polynomial resultant of solving \eqref{Eq: Imax_SOCsur}. The maximum discharge current can be obtained by solving the following quadratic expression:
\begin{IEEEeqnarray}{rCl}
0&=&a_{dis} \cdot \big(\overline{I}^\text{dis}_0\big)^2 + b_{dis}\cdot \overline{I}^\text{dis}_0 + c_{dis}, 
\IEEEeqnarraynumspace 
\label{Eq: Discharge Limits}\\
\noalign{\noindent with
\vspace{2\jot}}
a_{dis} &=& - R_{dif,elec} \cdot \eta_{c,i}, 
\IEEEeqnarraynumspace \IEEEyessubnumber*
\label{Eq: a_dis}\\
b_{dis} &=& - R_{dif,elec}[\eta_{c,0} + \eta_{c,T} \cdot T],
\IEEEeqnarraynumspace
\label{Eq: b_dis}\\
c_{dis} &=& SOC.
\IEEEeqnarraynumspace
\label{Eq: c_{dis}}
\end{IEEEeqnarray}

\begin{figure}[h]
	\centering
	\subfigure[Discharge current]{
 		\includegraphics[width=0.4655\linewidth]{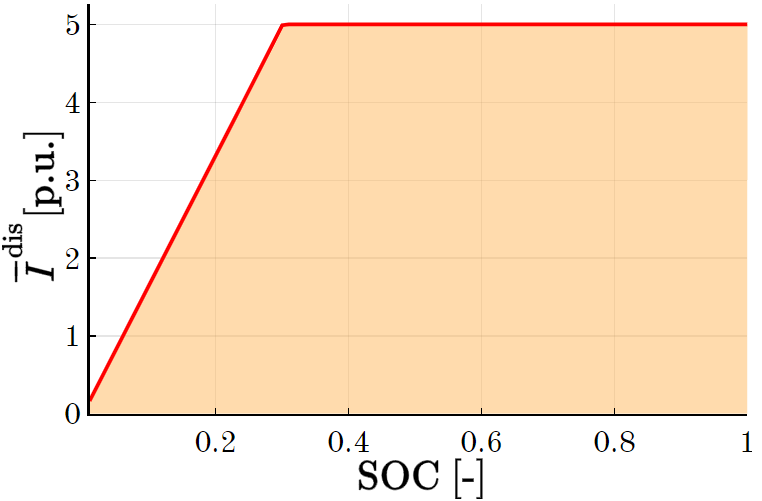}
  		\label{Fig: i^max_dis}}
	\subfigure[Charge current]{
  		\includegraphics[width=0.4655\linewidth]{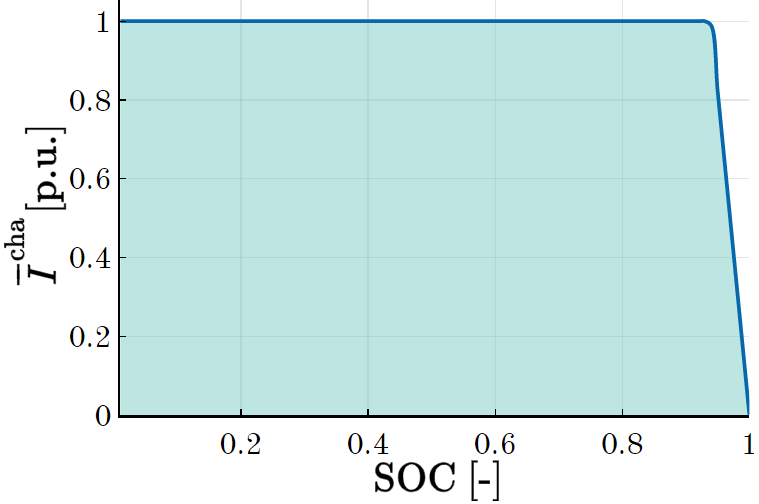}
  		\label{Fig: i^max_cha}}
  		\caption{Maximum current normalized to the battery capacity, as a function of the SOC. Shadow areas indicate feasible operation.}
\end{figure}

Analogously, by bounding the $SOC_{sur}$ in \eqref{Eq: SOC_sur bounded} by its upper limit, the maximum charging power can be obtained with the following quadratic expressions:
\begin{IEEEeqnarray}{rCl}
0 &=& a_{cha} \cdot \big(\overline{I}^\text{cha}_0\big)^2 + b_{cha}\cdot \overline{I}^\text{cha}_0 + c_{cha}, 
\IEEEeqnarraynumspace
\label{Eq: Charge Limits}\\
\noalign{\noindent with
\vspace{2\jot}}
a_{cha} &=& - R_{dif,elec} \cdot \eta_{c,i}, 
\IEEEeqnarraynumspace \IEEEyessubnumber*
\label{Eq: a_cha}\\
b_{cha} &=& - R_{dif,elec}[\eta_{c,0} + \eta_{c,T} \cdot T],
\IEEEeqnarraynumspace
\label{Eq: b_cha}\\
c_{cha} &=& SOC-1.
\IEEEeqnarraynumspace
\label{Eq: c_cha}
\end{IEEEeqnarray}
\color{black}
The current limits $\overline{I}^\text{cha/dis}_0$ refer to the limitations in the charge transfer process. To consider the manufacturer limits which prevent cell damage, the parameter $\overline{I}^\text{cha/dis}_{c-rate}$ is introduced. Thus, the maximum permissible current for a battery can be calculated based on $\overline{I}^\text{cha/dis}_{0}$ and $\overline{I}^\text{cha/dis}_{c-rate}$ by:
\begin{IEEEeqnarray}{C}
\overline{I}^\text{cha/dis} = min\{\overline{I}^\text{cha/dis}_0,\overline{I}^\text{cha/dis}_{c-rate}\}
\label{Eq: I Limits}
\end{IEEEeqnarray}
From expressions \eqref{Eq: Discharge Limits}--\eqref{Eq: I Limits}, it is now possible to calculate the maximum permissible current as a function of the energy stored. 
The maximum permissible C-rates for the discharge and charge current, $\overline{I}^\text{dis/cha}_{c-rate}$, have been set to $5C$\footnote{A C-rate of $5C$ indicates 5 times the nominal current for discharging according to typical manufacturer specifications.} and $1C$ to match the typical manufacturer limits used in \cite{Berrueta2018}. Accordingly, we have plotted the maximum feasible working points for discharging and charging currents vs. the SOC in Figs \ref{Fig: i^max_dis} and \ref{Fig: i^max_cha}.

As shown in Fig. \ref{Fig: i^max_dis} for low $SOC$, the maximum current for discharging decreases.
At low $SOC$, the internal battery resistance, $R_{tot}= R_{ohm} + R_{ct} + R_{dif,mem}$, increases considerably, decreasing the equivalent voltage, $v_{eq}$, below zero, which indicates an erroneous sense of battery depletion. This behavior corresponds to the voltage cut-off in the battery cells, which would result in a battery shut down by the management system \cite{Dong2018}.

The charging current limits deviate from the $1C$ rating for higher values of $SOC$, which is greater than 0.93, Fig. \ref{Fig: i^max_cha}. The physical limitations for cell charging correspond to a greater rate of voltage rise, when compared with the rate of charge absorption. This results in the saturation of electrochemical cells and an increase in stresses within the cells.

The expressions for calculating \textit{discharging power limits} \eqref{Eq: Calculating PdLim}, $\overline{P}^\text{dis}$, and \textit{charging power limits} \eqref{Eq: Calculating PcLim}, $\overline{P}^\text{cha}$, are derived using the circuit-equivalent battery model and currents limits.
\begin{IEEEeqnarray}{C} \label{Eq: QLims}
	\overline{P}^\text{dis} = v_{eq}\cdot\overline{I}^{dis}-\big(\overline{I}^{dis}\big)^2\cdot R_{tot}
		\IEEEyesnumber\IEEEyessubnumber*\label{Eq: Calculating PdLim}\\
	\overline{P}^\text{cha} = v_{eq}\cdot\overline{I}^{cha}+\big(\overline{I}^{cha}\big)^2\cdot R_{tot}
	\label{Eq: Calculating PcLim}
\end{IEEEeqnarray}
\subsection{Charging and Discharging Battery Efficiencies}
\begin{figure}[h!]
	\centering
	\subfigure[Discharge efficiency]{
  		\includegraphics[width=0.465\linewidth]{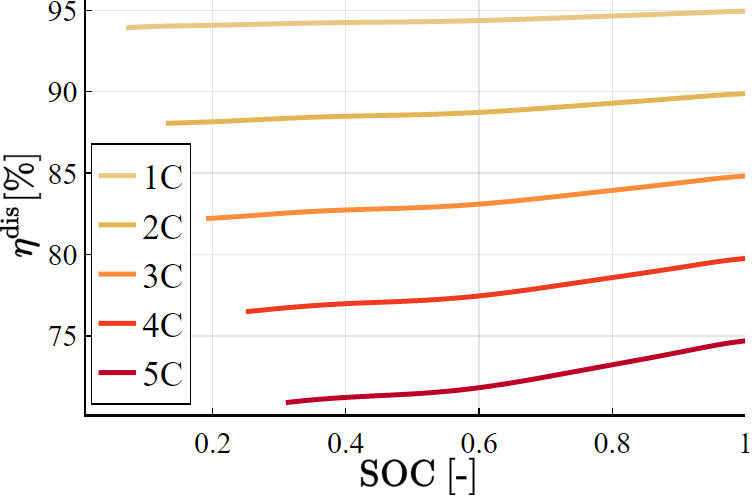}
  		\label{Fig: Dis Efficiency}}
	\subfigure[Charge efficiency]{
  		\includegraphics[width=0.465\linewidth]{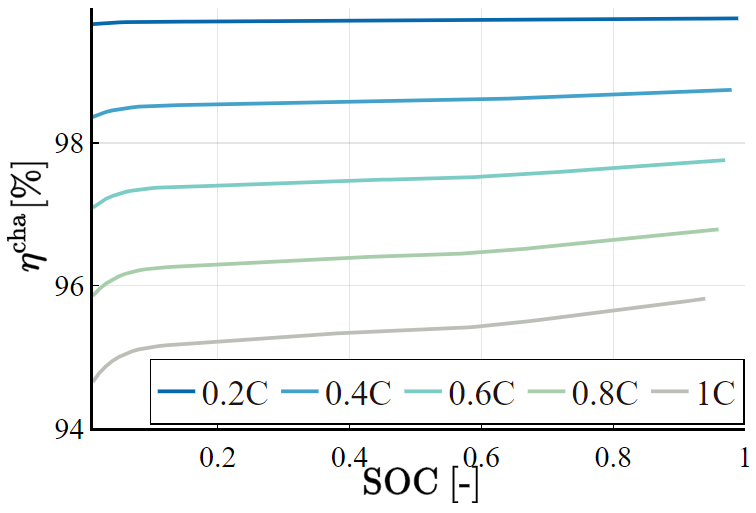}
  		\label{Fig: Cha Efficiency}}
        \caption{Discharging and charging efficiencies vs. the SOC and discharging and charging power}
\end{figure}

To characterize battery usage on an operation optimization model, it is necessary to derive an expression for the battery's performance, both for charging and discharging regimes.
The battery \textit{discharging efficiency} is given by \eqref{Eq: EffDis1}. The symbol $p^{\text{dis}}$ denotes the power discharged to the electric grid, and so $p^{\text{dis}} = v_{}\cdot i$. The symbol $p^\text{out}$ denotes outgoing power from the battery cells, $p^\text{out} =v_{eq}\cdot i$. Applying Kirchoff's Voltage Law on the circuit in Fig. \ref{Fig: Circuit}, we can derive the discharging efficiency of the battery.
\begin{IEEEeqnarray}{rClCl}
\eta^\text{dis} &=& \frac{p^\text{dis}}{p^\text{out}}&=& 1 - \frac{i\cdot R_{tot}}{v_{eq}}
\label{Eq: EffDis1}
\end{IEEEeqnarray}
%
%
Similar to discharging efficiency, we can derive a \textit{charging efficiency} expression as follows 
\begin{IEEEeqnarray}{rCl}
\eta^\text{cha} &=& \frac{p^\text{in}}{p^\text{cha}} = \frac{v_{eq}}{v_{eq}+i \cdot R_{tot}}
\label{Eq: EffCha1}
\end{IEEEeqnarray}
As seen in Fig. \ref{Fig: Dis Efficiency} and \ref{Fig: Cha Efficiency}, the efficiency of a battery improves for higher $SOC$ and C-rates closer to $1C$. This result corresponds to previous experimental analyses performed in \cite{Rampazzo2018,Schimpe2018,Sakti2017}. The discharging efficiency lowers for higher current values and low $SOC$, dropping as much as $33\%$ from its maximum value. The charging efficiency presents a similar behavior, with a smaller efficiency drop correspondent to the smaller operating region, $0-1C$.
\subsection{Non-Linear Battery Model}
By considering the aforementioned charging and discharging power limits and efficiencies, we can derive a new detailed battery model that is similar to conventional methodologies \cite{Pozo2014}. The approach is presented in \ModelOne. Equations \eqref{Eq:PdLim} and \eqref{Eq:PcLim} indicate the discharging and charging power limits. Contrary to conventional models, the limits are SOC-dependent. Equation \eqref{Eq. BalanceBattery} denotes the energy balance along the time steps. The energy stored in the battery at time $t$ is calculated as the sum of the energy stored in the previous time step $t-1$ and the energy entering the battery cell in the previous time-step. It is given by the product of $p^\text{cha}_{t-1} \eta_{t-1}\Delta$ minus the energy exiting the cell, which is represented by the term $p^\text{dis}_{t-1} \dfrac{1}{\eta_{t-1}^\text{dis}}\Delta$. The parameter $\Delta$ is the size of time steps as an hourly fraction; employed to transform the use of power into energy. Battery energy capacity limits are described by \eqref{Eq: Battery Capacity}. The battery state of charge, $SOC_t$, in p.u., is calculated as a function of the stored energy, $e_t$, by \eqref{Eq: SOC M1}. Finally, efficiencies and power limits are included in \eqref{Eq. others}.

\begin{model}[h]
\caption{NLP Li-ion battery model}
\vspace{-0.3cm} 
\label{mod:NLPModel}
\begin{IEEEeqnarray}{lCl}
\mbox{\bf Variables:} \nonumber \\
p_t^\text{dis}, p_t^\text{cha}  &-&  \mbox{discharging and charging power} \nonumber \\
\overline{P}_t^\text{dis}, \overline{P}_t^\text{cha} &-&   \mbox{maximum discharging and charging power}  \nonumber \qquad \quad\\
\eta_t^\text{dis}, \eta_t^\text{cha} &-& \mbox{discharging and charging efficiency}  \nonumber\\
e_t, SOC_t &-& \mbox{battery energy level (absolute and relative values)}  \nonumber 
\end{IEEEeqnarray}
\mbox{\bf Constraints:} \nonumber
\begin{subequations}  \label{eq.NLPModel}
\begin{IEEEeqnarray}{l"l}
0 \leq p^\text{dis}_t \leq \overline{P}^\text{dis}_t,  &\forall t \label{Eq:PdLim} \IEEEyesnumber\\
0 \leq p^\text{cha}_t \leq \overline{P}^\text{cha}_t,  &\forall t \label{Eq:PcLim} \\
e_t = e_{t-1} + p^\text{cha}_{t-1} \eta_{t-1}^\text{cha} \Delta 
- p^\text{dis}_{t-1} \frac{1}{\eta_{t-1}^\text{dis}}\Delta , \quad   &\forall t \label{Eq. BalanceBattery} \IEEEeqnarraynumspace\\
0 \leq e_t \leq \overline{E}, \quad &\forall t \label{Eq: Battery Capacity} \\
SOC_t =e_t/\overline{E}, \quad &\forall t \label{Eq: SOC M1} \\
\eqref{Eq: QLims}-\eqref{Eq: EffCha1}  \label{Eq. others}
\vspace{-2\jot}
\end{IEEEeqnarray}
\end{subequations}
\end{model}

\vspace{-2\jot}\section{Linear Reformulation Approach} \label{Section: Convex Combination}
In the aforementioned non-linear \ModelOne, non-convexities arise from discharging and charging power limits and efficiencies \eqref{Eq. others}, including bilinear products in the energy balance constraint  \eqref{Eq. BalanceBattery}. In this section, we propose a linear approach through a convex envelope for the characterization of battery charge and discharge.

To handle the bilinear products in the battery energy balance equation, we add two new variables, incoming and outgoing power from battery cells, $p^\text{in}$ and $p^\text{out}$, respectively. They are introduced in equations \eqref{Eq: EffDis1} and \eqref{Eq: EffCha1} and given by:
\begin{IEEEeqnarray}{C}
p^\text{out} = p^\text{dis}\frac{1}{\eta^\text{dis}}\label{Eq: Pin/out} \IEEEyesnumber\IEEEyessubnumber*\label{Eq: Pout}\\ 
p^\text{in} = p^\text{cha}\eta^\text{cha} \label{Eq: Pin}
\end{IEEEeqnarray}
Expressions in \eqref{Eq: Pin/out} allow us to define the energy balance equation as an affine function of $p^{out}$ and $p^{in}$
\begin{IEEEeqnarray}{C}
e_t = e_{t-1} + p^\text{in}_{t-1}\Delta - p^\text{out}_{t-1}\Delta.  \label{Eq: EBalance Affine}
\end{IEEEeqnarray}
This substitution can be done without the need for evaluating the bilinear products of \eqref{Eq: Pin/out} because $\eta^\text{out}$ is dependent on the power provided by the battery, $p^\text{dis}$, and the $SOC$, e.g. a function of $i$ and $v_{eq}$, see \eqref{Eq: EffDis1}. Therefore, the values of $p^\text{out}$ can be obtained in terms of $p^\text{dis}$ and $SOC$, which are the variables of interest in operation models implementing electric energy storage through batteries. 

\begin{figure}[h]
    \centering
    \subfigure[Discharging power]{
    \includegraphics[width=9cm,height=5cm,keepaspectratio]{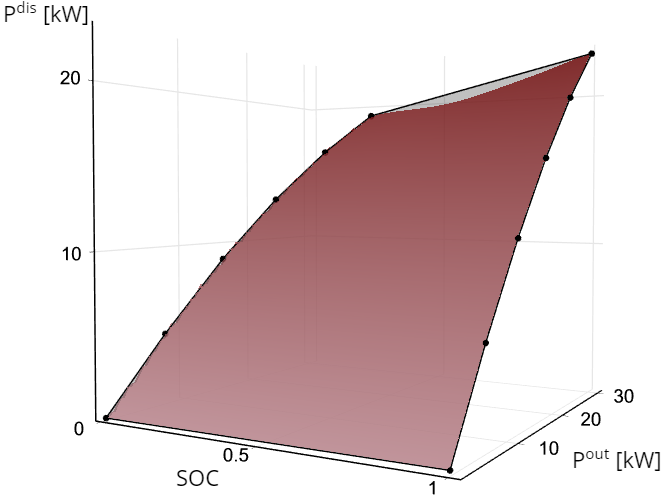}
    \label{Fig: P^out}}\\
    \subfigure[Charging power]{    \includegraphics[width=9cm,height=5cm,keepaspectratio]{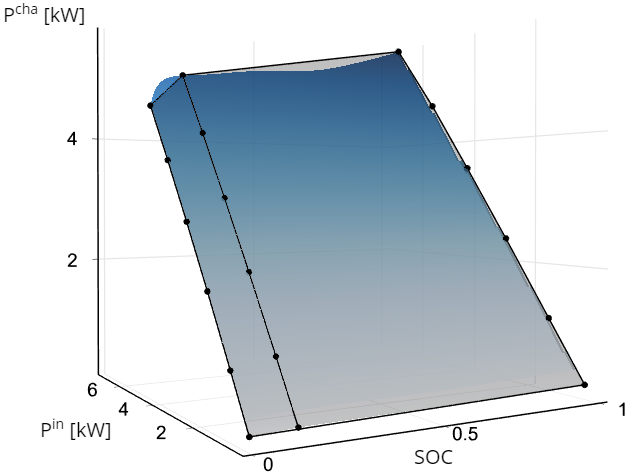}
    \label{Fig: P^in}}
    \caption{Operating region of Li-ion battery in variable space of (a) $[p^\text{dis}, SOC, p^\text{out}]^{\top}$, and (b) $[p^\text{cha}, SOC, p^\text{in}]^{\top}$. The curve indicates non-linear dependence, black dots denote sampled points, and the lines between the sampled points define the convex envelope of the sampled points. \label{fig: P-surfaces}}
\end{figure}

The outgoing power from the battery cells, $p^\text{out}$, can be approximated by a convex combination of sampling points. That is, we can construct a polyhedral envelope of the $p^\text{out}$ by sampling from the model given in Section \ref{Sect: Battery Physics}, or alternatively, by experiments like in \cite{Berrueta2018}. Therefore, for every tuple $[p^\text{dis}, SOC, p^\text{out}]^{\top}$, we draw $J$ samples through simulations, represented by $[\widehat{P}^\text{dis}_j, \widehat{SOC}_j,\widehat{P}^\text{out}_j]^{\top}$. The linear convex envelope is formulated in \eqref{Eq: Pout Convex Comb}. 
A 3-dimensional representation of $p^\text{dis}$ vs. $SOC$ and $p^\text{out}$ is presented in Fig. \ref{Fig: P^out}, where the solid areas indicate the simulated values based on Eq. \eqref{Eq: Pout}. The sampled points $j \in J$ are highlighted as black dots, and the convex envelope connecting the sampled points is shown by the connecting lines between the points. As it can be observed, the proposed approach based on the convex envelope for a feasible set points is very close to the non-linear mathematical definition. 
\begin{IEEEeqnarray}{lCll} \label{Eq: Pout Convex Comb}
p^\text{out} &=& \sum_j \widehat{P}^\text{out}_{j} \cdot x_j \IEEEyesnumber*\IEEEyessubnumber*\\
p^\text{dis}  &=& \sum_j \widehat{P}^\text{dis}_{j} \cdot x_j\\
SOC &=& \sum_j \widehat{SOC}_{j} \cdot x_j \label{Eq: SOC_j}\\
1 &=& \sum_j x_j \label{Eq: x_j}\\
0 &\le& x_j, &\qquad \forall j \in J
\end{IEEEeqnarray}
Analogously, $p^\text{in}$ can be approached by the convex envelope defined in \eqref{Eq: Pin Convex Comb}. Similarly, it is also compared with the non-linear definition in Fig. \ref{Fig: P^in}.
\begin{IEEEeqnarray}{lCll} \label{Eq: Pin Convex Comb}
p^\text{in} &=& \sum_k \widehat{P}^\text{in}_{k}y_k \IEEEyesnumber*\IEEEyessubnumber*\\
p^\text{cha}  &=& \sum_k \widehat{P}^\text{cha}_{k}y_k\\
SOC &=& \sum_k \widehat{SOC}_{k}y_k \label{Eq: SOC_k}\\
1 &=& \sum_k y_k \label{Eq: y_k}\\
0 &\le& y_{k}, &\qquad \forall k \in K
\end{IEEEeqnarray}

The expressions for battery characterization are given as a function of the $SOC$ as a consequence of the dependence of the $v_{eq}$ and power limits on it, including the computational advantages of employing a normalized parameter.
The resultant linear model for battery characterization is presented in \ModelTwo.

\begin{model}[!h]
\caption{Linear Li-ion battery model}
\vspace{-0.3cm} 
\label{Mod: CCModel}
\begin{subequations}  \label{Eq: CCModel}
\begin{IEEEeqnarray}{lCl}
\mbox{\bf Variables:} \nonumber \\
e_t, SOC_t &-& \mbox{battery energy level (absolute and relative values)}  \nonumber \\
p_t^\text{dis}, p_t^\text{cha}  &-&\mbox{discharging and charging power}  \nonumber \\
p_t^\text{out}, p_t^\text{in}  &-&\mbox{power outgoing and incoming at the cells}  \nonumber \\
x_{jt}, y_{kt}   &-&\mbox{variables related to the sample sets $J$ and $K$}  \nonumber
\end{IEEEeqnarray}
\mbox{\bf Constraints:} \nonumber
\begin{IEEEeqnarray}{lCll}
e_t &=& e_{t-1} + \big(p^\text{in}_{t-1} - p^\text{out}_{t-1}\big)\Delta  \label{Eq: EBalance Affine Model2}, &\forall t \\
p^\text{out}_t &=& \sum_j \widehat{P}^\text{out}_{jt}x_{jt}, &\forall t \\
p^\text{dis}_t  &=& \sum_j \widehat{P}^\text{dis}_{jt}x_{jt}, &\forall t \\
p^\text{in} &=& \sum_k \widehat{P}^\text{in}_{kt}y_{kt}, &\forall t \\
p^\text{cha}  &=& \sum_k \widehat{P}^\text{cha}_{kt}y_{kt}, &\forall t \\
SOC_t &=& \sum_j \widehat{SOC}_{jt}x_{jt} + \sum_k \widehat{SOC}_{kt}y_{kt}, \quad \label{Eq: SOC_jk} \qquad &\forall t \\
SOC_t &=& e_t/\overline{E},  &\forall t \label{Eq: Battery SOC_M2}\\
1 &=& \sum_j x_{jt}, &\forall t \\
1 &=& \sum_k y_{kt}, &\forall t \\
0 &\le& x_{jt}, &\forall j, t \\
0 &\le& y_{kt}, &\forall k, t \IEEEeqnarraynumspace
\vspace{-2\jot}
\end{IEEEeqnarray}
\end{subequations}
\end{model}

The $SOC$ is characterized by \eqref{Eq: SOC_j} and \eqref{Eq: SOC_k}. If both expressions were jointly considered, the battery would appear to be charging and discharging at the same time. This corresponds to the fact that to guarantee an approximation through a convex combination of sampling points, constraints \eqref{Eq: x_j} and \eqref{Eq: y_k} require that at least one $x_j$ and one $y_j$ be greater than zero, simultaneously making $p^\text{dis}$ and $p^\text{cha}$ non zero. In \ModelOne{}, this does not occur because the energy balance, \eqref{Eq. BalanceBattery}, uses efficiencies $\eta^\text{dis}$ and $\eta^\text{cha}$ that are lower than one. Therefore, simultaneously charging and discharging would go against the economic objective of minimizing the cost of power system operations because energy would be lost during the imperfect (and simultaneous) charge and discharge processes.\footnote{This behavior is ensured for a majority of power system applications where the power balance constraints can be satisfied within the technical limits of the generators and demand, i.e., when the demand can be fulfilled without recurring load shedding or generation curtailment.} Based on convex combination constraints and the economic use of the battery, constraints \eqref{Eq: SOC_j} and \eqref{Eq: SOC_k} can be combined through a summation in \eqref{Eq: SOC_jk}. For this constraint to allow a discernment of the charging and discharging processes without the introduction of binary variables, an additional condition is introduced: each sampling set must have at least two sampling points equal to $[0, 0, 0]^{\top}$ and $[1, 0, 0]^{\top}$; which respectively represent the cases when the battery is not active $p=0$, but is fully discharged or fully charged. Consequently, for a discharge $p^\text{dis}>0$, $x_{j_0} = 0$ and then $p^\text{cha} = 0$, since $y_{j_0} = 1$; an analogous relationship would follow for the charging cycle.

For the considered discharging convex approximation by our proposed method we have a maximum approximation error of 9.03\%, a mean error of 1.21\%, and a standard deviation of 1.39\%, when 14 sampling points are considered. 
For the charging process  sampled with 20 points, we have a maximum approximation error is of 1.12\%, with a mean error of 0.22\% and a standard deviation of 0.18\%. 
The greater approximation error for the discharging curve can be explained by the greater changes in current that the sampling points in the y-axis (6 points for both the charge and discharge) must approximate, when compared to the charging range.

\section{Network-Constrained Economic Dispatch with Energy Storage Devices} \label{Section: Economic Dispatch}
A conventional economic dispatch, based on a lossless DC approximation, with linear costs is modeled by \ModelED{}.
The scheduled cost of energy generation is given by \eqref{Eq: sum Gen Cost}. 
Equation \eqref{Eq: P Balance} represents the power balance at every node $n$ of the system for every time step. The power entering the node from each connected line, $f_{lt}$, is equal to the nodal demand, $P^\text{D}_{nt}$, minus the power generated at the node, $p_{gt}$, minus the discharging power of the battery connected to this node, $p^\text{dis}_{st}$, plus its charging power, $p^\text{cha}_{st}$. The power flowing in line $l$ is modeled by \eqref{Eq: Line Flow} using line-to-node incidence matrix, $A_{nl}$. 
Equations \eqref{Eq: Gen Limits}, \eqref{Eq: Line Flow Limits}, and \eqref{Eq: Angle Limits} establish the technical limits of the generators $g$, power lines $l$ and voltage angle at node $n$, respectively. The reference voltage angle is set by \eqref{Eq: Ref Angle}. Battery energy level at the end of a dispatch horizon is set to be equal to the initial battery energy level \eqref{Eq: Battery Cycle}.\vspace{-2\jot}
\begin{model}[h]
\caption{Network-Constrained Economic Dispatch with EES}
\vspace{-0.3cm} 
\label{Mod: EDModel}
\vspace{4.5\jot}
\textbf{Variables:}
\begin{IEEEdescription}[\IEEEusemathlabelsep\IEEEsetlabelwidth{$\delta_{nt}$}]
\item[$p_{gt}$] generated power by $g$ during $t$
\item[$f_{lt}$]  line power flow  through $l$ on $t$
\item[$\delta_{nt}$]  voltage phase angle at $n$ on $t$
\end{IEEEdescription}
\mbox{\bf Objective:} \nonumber\vspace{-2\jot}
\begin{IEEEeqnarray}{l}
\minimize \qquad \sum\limits_{g,t} C_g p_{g,t} \Delta
    \label{Eq: sum Gen Cost}\vspace{-2\jot}
\end{IEEEeqnarray}
\vspace{-2\jot}\mbox{\bf Constraints:} \nonumber
\begin{IEEEeqnarray}{lr}
{\sum\limits_{l}} A_{nl}f_{lt} {=} P_{nt}^\text{D} {-}{\sum\limits_{g}}\Omega_{gn}{\cdot}p_{gt} {-} \sum_s\Gamma_{sn}[p_{st}^\text{dis} {-} p_{st}^\text{cha}], & \forall n,t \IEEEyesnumber\IEEEyessubnumber*\IEEEeqnarraynumspace \label{Eq: P Balance}\\
f_{lt} =\frac{1}{X_l} \sum\limits_{n}A_{nl} \cdot \delta_{nt},  & \forall l \IEEEeqnarraynumspace \label{Eq: Line Flow}\\
 \underline{P}_g \leq p_{gt}  \leq \overline{P}_g, & \forall i,t \IEEEeqnarraynumspace\label{Eq: Gen Limits}\\
-\overline{F}_l \leq f_{lt}  \leq \overline{F}_l, & \forall l
    \IEEEeqnarraynumspace\label{Eq: Line Flow Limits}\\
\underline{\delta}_n \leq \delta_{n,t} \leq \overline{\delta}_n, & \forall n \IEEEeqnarraynumspace \label{Eq: Angle Limits}\\
\delta_{n=n_0,t} = 0, & \forall t \IEEEeqnarraynumspace \label{Eq: Ref Angle}\\
e_{t=1} = e_{t=|T|} \label{Eq: Battery Cycle} \\
\text{Battery model (\ModelTwo): } \eqref{Eq: CCModel}.
\vspace{-2\jot}
\end{IEEEeqnarray}
\end{model}
\section{Test Case} \label{Section: Test Case}
We evaluated the effect of the proposed characterization on battery operation for electric economic dispatch. For this purpose, four model approaches were compared:
\begin{itemize}
    \item \CaseA: the storage systems is modeled as described in \ModelOne{}, resulting a non-linear non-convex (NLP) problem.
    \item \CaseB{}: an ideal battery was considered, i.e., \ModelOne{} is employed, but the parameters $\overline{P}^\text{cha}$, $\overline{P}^\text{dis}$, $\eta^\text{cha}_t$, and $\eta^\text{dis}_t$ were assumed to be constant.
    \item \CaseC: the battery is modeled as described in \ModelTwo, employing a convex piecewise linearization of its characteristic curve.
    \item \CaseD: for the representation of the non-convex nature of the characteristic curves, Fig. \ref{fig: P-surfaces}, a piecewise linearization using binary variables is employed \cite{DAmbrosio2010a}. The resultant model is a mixed-integer linear programming (MILP) one.
\end{itemize}
The computational tests were performed using a modified version of the IEEE Reliability Test System (IEEE RTS) \cite{IEEE-RTS}. 
The generator technical data, the line parameters, the load profile, including the battery parameters, are given in the online dataset \cite{GonzalezC_DataBatt}.
The optimization horizon was set to 24 hours, with 10-minute intervals. The load data was scaled to the IEEE RTS system based on the demand data from the Iberian Electricity Market on June 20, 2018 \cite{REE}.  The simulations are performed using the modeling software Julia 0.6.4 \cite{Bezanson2017}, JuMP 0.18 \cite{Lubin2015}, and Gurobi 7.5.1. \cite{gurobi} and IPOpt \cite{IpOpt} as solvers. A summary of the characteristics and results of the evaluated cases is provided in Table \ref{Tab: Results}.
\subsection{Results}
\subsubsection{\CaseA - Non-linear programming model}
The general non-linear programming (NLP) model, \ModelOne{}, has been evaluated to provide a reference for the battery operation. The scheduled battery operation is presented for both charge and discharge in Fig. \ref{Fig: Cha_Time_NLP} and Fig. \ref{Fig: Dis_Time_NLP}, respectively. The real maximum charging and discharging power limits have been calculated based on \eqref{Eq: Calculating PdLim}--\eqref{Eq: Calculating PcLim} and represented by the red dashed line in the figures. Battery usage follows charging cycles for periods of lower demand, to subsequently discharge during higher load request. Power limits for both charge and discharge change throughout the day as a function of the stored energy, as described in Section \ref{Sec: Limits}.
\subsubsection{\CaseB -- Ideal battery model}
\begin{figure}[!h]
	\centering
	\subfigure[\CaseA{}: charging power vs. time]{
  		\includegraphics[width=0.8\linewidth]{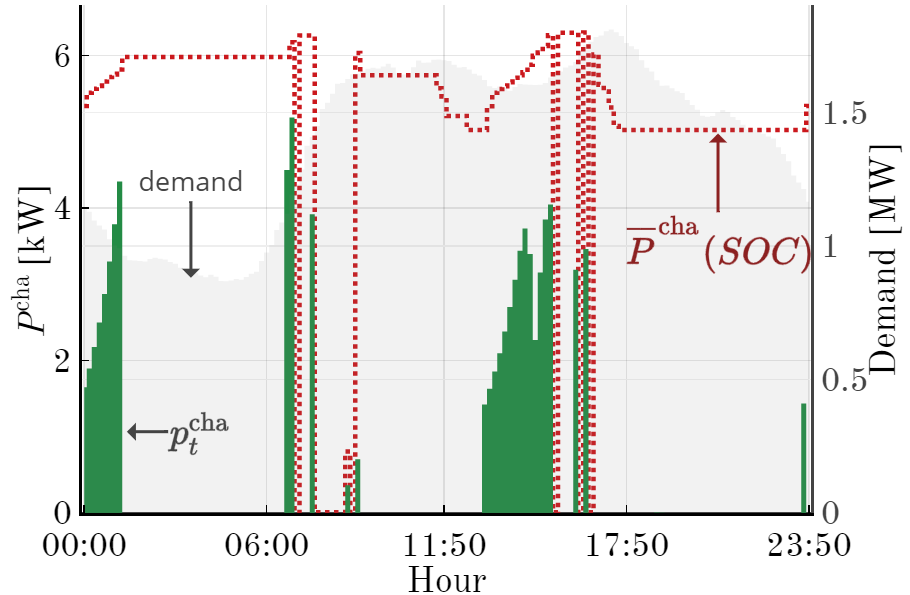}
  		\label{Fig: Cha_Time_NLP}}
	\subfigure[\CaseA{}: discharging power vs. time]{
  		\includegraphics[width=0.8\linewidth]{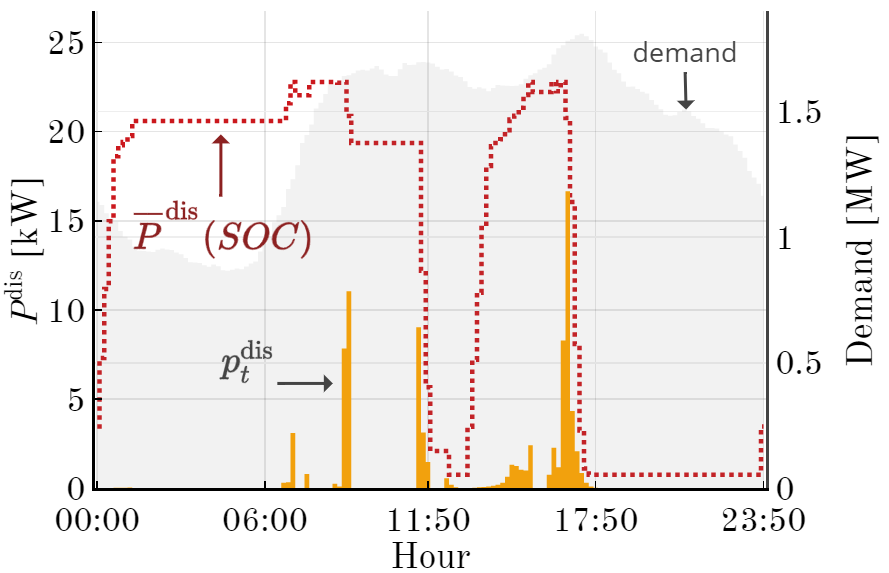}
  		\label{Fig: Dis_Time_NLP}}\\
	\subfigure[\CaseB{}: charging power vs. time]{
  		\includegraphics[width=0.8\linewidth]{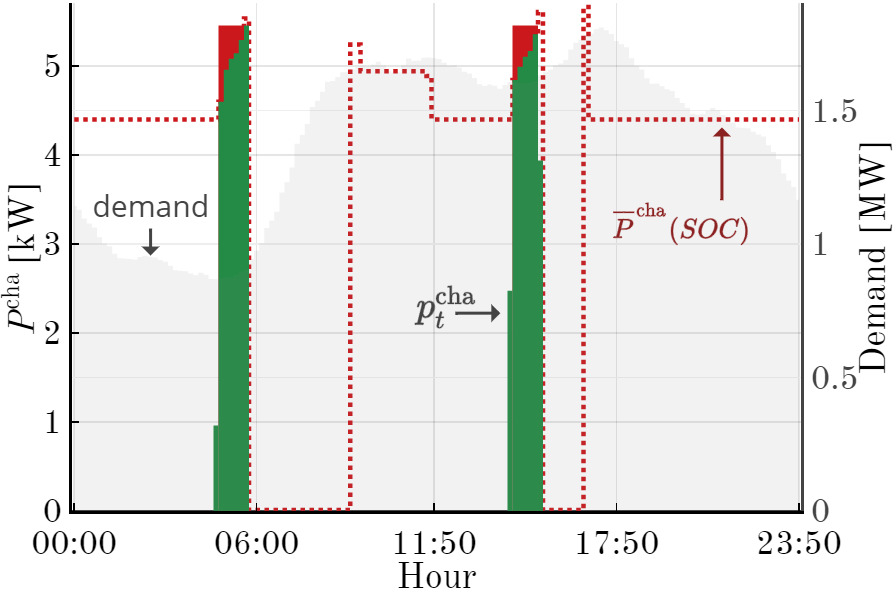}
  		\label{Fig: Cha_Time_LP}}
	\subfigure[\CaseB{}: discharging power vs. time]{
  		\includegraphics[width=0.8\linewidth]{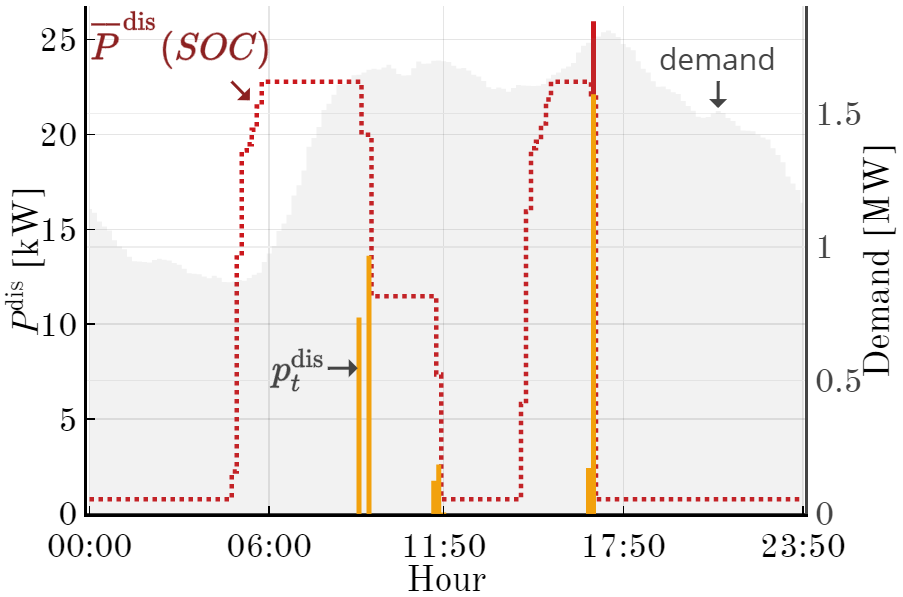}
  		\label{Fig: Dis_Time_LP}}\\
	\caption{Scheduling of the battery charge and discharge processes for (a)-(b) \CaseA{}, and (c)-(d) \CaseB{}. The dotted line denotes the maximum power as a function of the $SOC$ based on \ModelOne{}. The red areas indicate infeasible operation.}
\end{figure}

In this case, the battery model is based on \cite{Pozo2014}. For doing so, we considered \ModelOne{}, where power limits and efficiencies have constant values. The current limits, \eqref{Eq: I Limits}, are set to 1C and 5C. The voltage at the battery terminals is set constant at the rated value of $133V$.
For the ideal battery model, the efficiencies are set to constant values, $\eta^\text{cha}_t=0.972$ and $\eta^\text{dis}_t=0.868$. The constant efficiencies were calculated as the mean of the values
given by \eqref{Eq: EffDis1} and \eqref{Eq: EffCha1}.\\
As a result of the economic dispatch defined in \ModelED{}, the battery was scheduled to charge from and discharge to the network as shown in Fig. \ref{Fig: Cha_Time_LP} and \ref{Fig: Dis_Time_LP}, respectively. As it can be seen, the scheduled operation of the ideal battery violates the calculated limits during peak charge and discharge.

\subsubsection{\CaseC -  Proposed linear battery model}
The scheduled operation of the proposed model is given in Fig. \ref{Fig: Cha_Time_PWL} and Fig. \ref{Fig: Dis_Time_PWL}, characterized by its higher use, i.e., during more time steps. For this battery model, the efficiencies depend on the state of charge and the power request, i.e., higher efficiency for higher SOC and lower power charge/discharge. Consequently, the battery use in this case must balance the power delivered with the efficiency and the system's marginal cost of generation. For this purpose, the battery is used at lower power levels, when compared to the ideal model with constant efficiency, to operate at higher efficiency. This change in scheduling can be observed in more time steps used for charge and discharge. Greater use of the non-ideal battery allows the objective value of this test case to be the same as that obtained for the ideal battery, even though the battery has a varying efficiency.
%
\begin{table}[]
\centering
\caption{Summary of optimization results}
\label{Tab: Results}
\begin{tabular}{@{}l|rrrr@{}}
\toprule
\multicolumn{1}{c}{}& \multicolumn{1}{c}{NLP} & \multicolumn{1}{c}{Ideal} & \multicolumn{1}{c}{MILP} & \multicolumn{1}{c}{PWL}  \\ \midrule
Objective value      & 57 305 & 57 312 & 57 307 & 57 308\\
$\Delta$ {[}\%{]}    & - & 0.01 & 0.01 & 0.01\\
Time {[}s{]}         & 231.3 & 1.2 & 143.3 & 3.4\\
Constraints          & 10 657 & 9 937 & 24 049 & 10 369                  \\
Continuous variables            & 18 576 & 11 088 & 20 592 & 16 272 \\
Binary variables     & - & - & 13 824 & - \\ \bottomrule
\end{tabular}
\end{table}
%
\subsubsection{\CaseD - Mixed-integer linear programming model}
\begin{figure}[!h]
	\centering
	\subfigure[\CaseC: charging power vs. time]{
  		\includegraphics[width=0.8\linewidth]{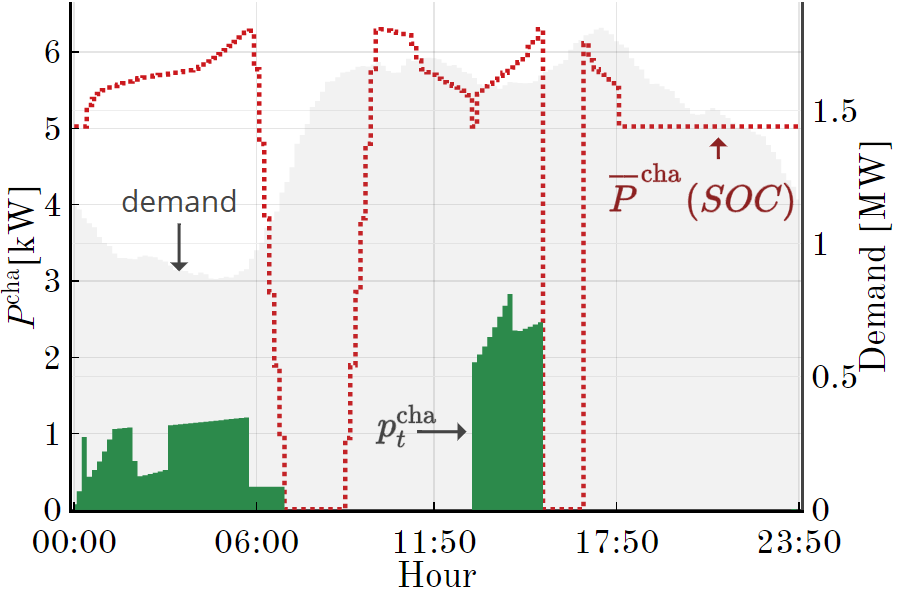}
  		\label{Fig: Cha_Time_PWL}}
	\subfigure[\CaseC: discharging power vs. time]{
  		\includegraphics[width=0.8\linewidth]{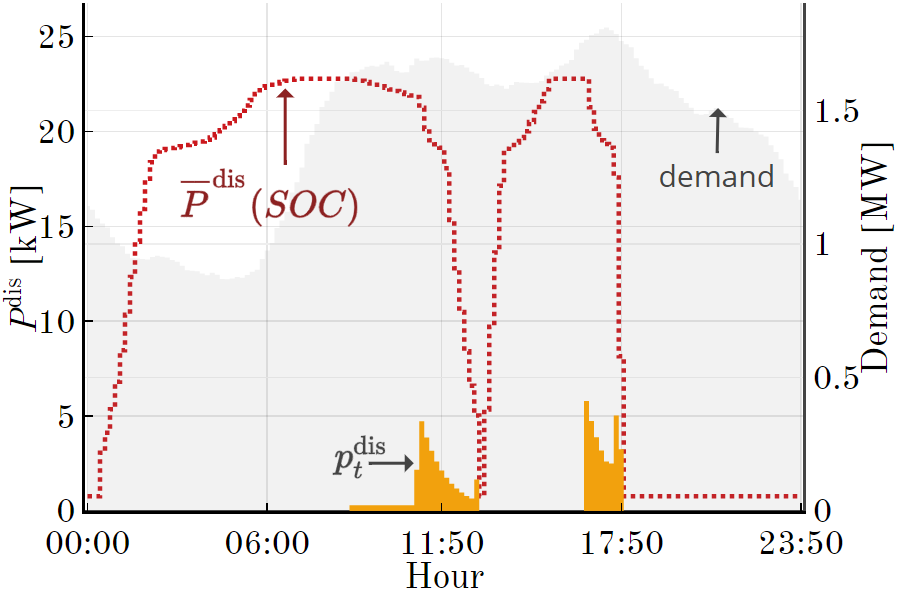}
  		\label{Fig: Dis_Time_PWL}}\\
	\subfigure[\CaseD: charging power vs. time]{
  		\includegraphics[width=0.8\linewidth]{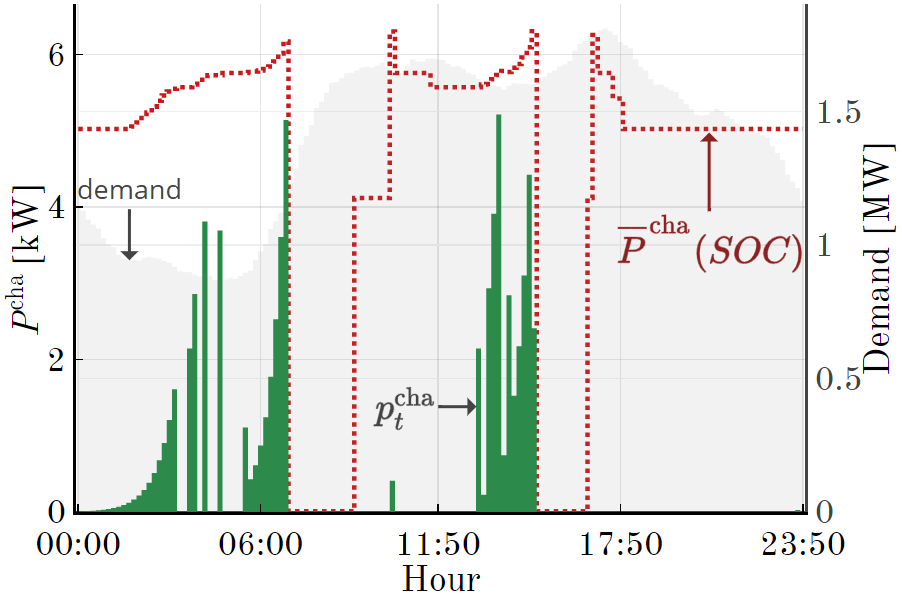}
  		\label{Fig: Cha_Time_MILP}}
	\subfigure[\CaseD: discharging power vs. time]{
  		\includegraphics[width=0.8\linewidth]{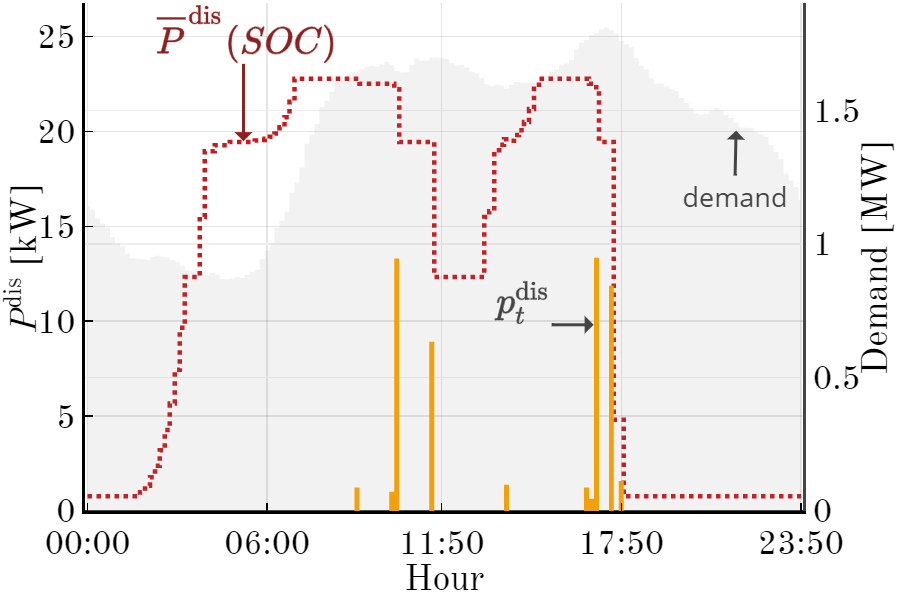}
  		\label{Fig: Dis_Time_MILP}}\\
	\caption{Scheduling of the battery charge and discharge processes for (a)-(b) \CaseC{}, and (c)-(d) \CaseD{}. The dotted line denotes the maximum power as a function of the $SOC$ based on \ModelOne{}. The red areas indicate infeasible operation.}
\end{figure}

With the aim of comparing the proposed convex model with one that captures the non convexity nature of the battery characteristic curves, a mixed-integer programming model (MILP) has been used to represent the non convex piecewise linearization of the curve through the triangle method \cite{DAmbrosio2010a}. The charging and discharging curves have been each represented with 15 sampling points, taking a base of 5 points for the SOC- and 3 for the $p^\text{out/cha}$-axis. The higher number of points in the SOC-axis allows a better representation of the power limits as a function of the stored energy.
\newline
The results of the MILP model are presented in Fig. \ref{Fig: Cha_Time_MILP} and Fig. \ref{Fig: Dis_Time_MILP}. This case employs the battery in two charging and discharging cycle with greater intensity, shorter time of use and higher requested power. The objective value resultant of the MILP model is the closest one, albeit by a small margin, to that provided by the complete non-convex formulation (NLP). Nonetheless, its computational cost is considerably higher than that of the proposed model, correspondent to the exponential increase of its solution time in relation to the number of binary variables. This makes it unsuitable for real life applications where the amount of batteries and size of the time horizon of interest is considerably larger.
%
%
\color{black}

\subsection{Reliability Model Assessment}

\begin{figure}[h]
	\centering
  	\includegraphics[width = 0.8\linewidth]{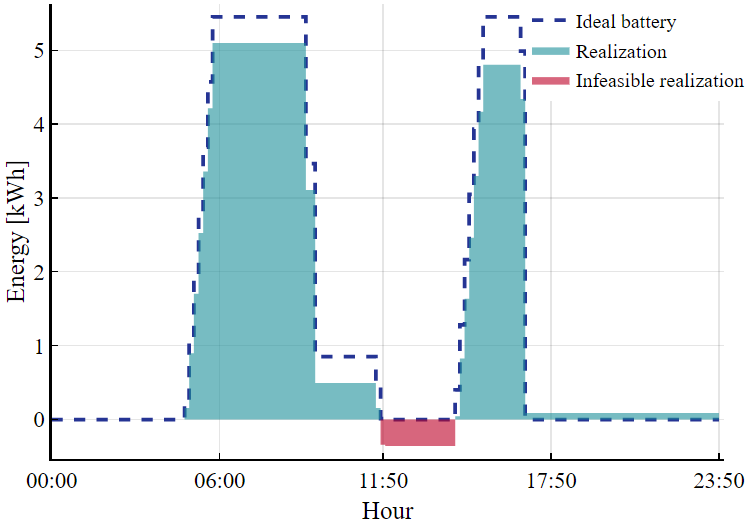}
	\caption{Analysis of deviations and unfeasible scheduling for an ideal battery\label{Fig: SOC_Realization}}
\end{figure}

The schedule of an ideal battery model for power charging and discharging is possible in unfeasible regions of battery operation. Therefore, it is expected that there will be situations in which the battery cannot provide the power/energy required by the schedule.
To calculate this mismatch, we introduce a reliability metric obtained as follows:
\begin{enumerate} 
  \item For the time steps in which $p^\text{cha}_t$ or $p^\text{dis}_t$ consider values outside of the regions defined by \eqref{Eq: Pout Convex Comb}--\eqref{Eq: Pin Convex Comb}, we set their values to be equal to the maximum power attainable for the given $SOC$.
  \item The realized (corrected) charging and discharging schedule for the battery is used for updating new energy levels, $e^{real}_t$, based on \eqref{Eq. BalanceBattery}.
  \item The energy imbalance/deviation is then calculated as the sum of the differences between the scheduled and the realized energy levels, as follows:
  \begin{equation}
  	\text{Imbalance} = \sum_t \big(e^{real}_t - e_t\big)
    \label{Eq: Deviation}
  \end{equation}
\end{enumerate}
Fig. \ref{Fig: SOC_Realization} presents the scheduled $SOC$ for an ideal battery and the realized levels resulting from this analysis. It can be observed that there exists not only a discrepancy between the scheduled and the realized values, but the battery would also reach negative energy levels, i.e., selling more energy than the available one. 
The energy deviation calculated by \eqref{Eq: Deviation} accounts for $12.2\%$ of the scheduled energy to be stored in the battery, resulting not only in a profit detriment for the owner but also on an overestimation of the system reliability that leads to a false sense of flexibility. 
%
\vspace{-4\jot}
\section{Conclusions} \label{Section: Conclusion}
This paper proposes a detailed model for battery characterization in optimization problems. The model is built based on a Li-ion battery, where the relationship between the state of charge, the charging and discharging efficiencies, and the power limits are described. 
The proposed model is based on a non-linear equivalent circuit model. A linear reformulation is then proposed based on a convex envelope for all feasible operation set-points. The linear reformulation is a sample-based approach but very close to the original non-convex and non-linear mathematical model.
The linear model enables the optimal management of a battery system without the use of binary variables. 

The proposed model has been integrated into an economic dispatch and tested against an ideal battery model, with constant power limits and efficiencies, that has been commonly used in the power systems literature. A case study for computational tests was performed based on the 24-bus IEEE-RTS system. The ideal model presents violations of the technical battery power limits, accounted for a $12\%$ deviation from the scheduled battery usage, highlighting the importance of such a detailed model to avoid wrong estimations of attainable flexibility and risking damaging the storage system.

Given the increasing need for flexibility and reliability in energy systems, the proposed linear model allows characterizing the operation and technical limits of electric battery systems through a computationally-efficient approach. The introduced battery characterization could be applied in combination with online state-of-charge and state-of-health estimators, to monitor the battery performance and the capacity fading in the battery. These systems could provide a continuously updating version of the battery capacity and performance, allowing to update the characteristic operation curves for charge and discharge presented in Section \ref{Section: Convex Combination}.
We advocate, then, for the use of more detailed battery models like the one proposed in this paper for consideration in existing power systems operation and planning problems that involve energy storage.\vspace{-4\jot} 

\ifCLASSOPTIONcaptionsoff
  \newpage
\fi



%
\bibliography{Ref_BMI}

\begin{thebibliography}{10}
\providecommand{\url}[1]{#1}
\csname url@samestyle\endcsname
\providecommand{\newblock}{\relax}
\providecommand{\bibinfo}[2]{#2}
\providecommand{\BIBentrySTDinterwordspacing}{\spaceskip=0pt\relax}
\providecommand{\BIBentryALTinterwordstretchfactor}{4}
\providecommand{\BIBentryALTinterwordspacing}{\spaceskip=\fontdimen2\font plus
\BIBentryALTinterwordstretchfactor\fontdimen3\font minus
  \fontdimen4\font\relax}
\providecommand{\BIBforeignlanguage}[2]{{%
\expandafter\ifx\csname l@#1\endcsname\relax
\typeout{** WARNING: IEEEtran.bst: No hyphenation pattern has been}%
\typeout{** loaded for the language `#1'. Using the pattern for}%
\typeout{** the default language instead.}%
\else
\language=\csname l@#1\endcsname
\fi
#2}}
\providecommand{\BIBdecl}{\relax}
\BIBdecl

\bibitem{Schimpe2018}
M.~Schimpe, M.~Naumann, N.~Truong, H.~C. Hesse, S.~Santhanagopalan, A.~Saxon,
  and A.~Jossen, ``Energy efficiency evaluation of a stationary lithium-ion
  battery container storage system via electro-thermal modeling and detailed
  component analysis,'' \emph{Applied Energy}, vol. 210, no. October 2017, pp.
  211--229, 2018.

\bibitem{TernaReport}
{Terna S.p.A.}, ``\BIBforeignlanguage{it}{{Public report year 2016: Pilot
  projects and testing of energy storage on energy-intensive batteries}},''
  {Terna S.p.A.}, Tech. Rep., Apr. 2017.

\bibitem{Lund2016a}
H.~Lund, P.~A. {\O}stergaard, D.~Connolly, I.~Ridjan, B.~V. Mathiesen,
  F.~Hvelplund, J.~Z. Thellufsen, and P.~Sorkn{\ae}s, ``Energy storage and
  smart energy systems,'' \emph{International Journal of Sustainable Energy
  Planning and Management}, vol.~11, pp. 3--14, 2016.

\bibitem{Fares2014}
R.~L. Fares and M.~E. Webber, ``A flexible model for economic operational
  management of grid battery energy storage,'' \emph{Energy}, vol.~78, pp.
  768--776, 2014.

\bibitem{Gunter2016}
N.~G{\"u}nter and A.~Marinopoulos, ``Energy storage for grid services and
  applications : {{Classification}} , market review , metrics , and methodology
  for evaluation of deployment cases,'' \emph{Journal of Energy Storage},
  vol.~8, pp. 226--234, 2016.

\bibitem{Uddin2018}
K.~Uddin, M.~Dubarry, and M.~B. Glick, ``The viability of vehicle-to-grid
  operations from a battery technology and policy perspective,'' \emph{Energy
  Policy}, vol. 113, no. August 2017, pp. 342--347, Feb. 2018.

\bibitem{Rappaport2017}
R.~D. Rappaport and J.~Miles, ``\BIBforeignlanguage{en}{Cloud energy storage
  for grid scale applications in the {{UK}}},''
  \emph{\BIBforeignlanguage{en}{Energy Policy}}, vol. 109, pp. 609--622, Oct.
  2017.

\bibitem{Rodriguez2017}
A.~Rodr{\'i}guez and G.~L. Plett, ``\BIBforeignlanguage{en}{Controls-oriented
  models of lithium-ion cells having blend electrodes. {{Part}} 2:
  {{Physics}}-based reduced-order models},''
  \emph{\BIBforeignlanguage{en}{Journal of Energy Storage}}, vol.~11, pp.
  219--236, Jun. 2017.

\bibitem{Ali2017}
D.~Ali, S.~Mukhopadhyay, H.~Rehman, and A.~Khurram,
  ``\BIBforeignlanguage{en}{{{UAS}} based {{Li}}-ion battery model parameters
  estimation},'' \emph{\BIBforeignlanguage{en}{Control Engineering Practice}},
  vol.~66, pp. 126--145, Sep. 2017.

\bibitem{Vagropoulos2013}
S.~I. Vagropoulos and A.~G. Bakirtzis, ``Optimal {{Bidding Strategy}} for
  {{Electric Vehicle Aggregators}} in {{Electricity Markets}},'' \emph{IEEE
  Transactions on Power Systems}, vol.~28, no.~4, pp. 4031--4041, Nov. 2013.

\bibitem{Pandzic2019}
H.~Pand{\v z}i{\'c} and V.~Bobanac, ``An {{Accurate Charging Model}} of
  {{Battery Energy Storage}},'' \emph{IEEE Transactions on Power Systems},
  vol.~34, no.~2, pp. 1416--1426, Mar. 2019.

\bibitem{Wang2016}
Y.~Wang, X.~Lin, and M.~Pedram, ``A {{Near}}-{{Optimal Model}}-{{Based Control
  Algorithm}} for {{Households Equipped With Residential Photovoltaic Power
  Generation}} and {{Energy Storage Systems}},'' \emph{IEEE Transactions on
  Sustainable Energy}, vol.~7, no.~1, pp. 77--86, Jan. 2016.

\bibitem{Morstyn2018}
T.~Morstyn, B.~Hredzak, R.~P. Aguilera, and V.~G. Agelidis, ``Model
  {{Predictive Control}} for {{Distributed Microgrid Battery Energy Storage
  Systems}},'' \emph{IEEE Transactions on Control Systems Technology}, vol.~26,
  no.~3, pp. 1107--1114, May 2018.

\bibitem{Berrueta2018}
A.~Berrueta, A.~Urtasun, A.~Urs{\'u}a, and P.~Sanchis, ``A comprehensive model
  for lithium-ion batteries: {{From}} the physical principles to an electrical
  model,'' \emph{Energy}, vol. 144, pp. 286--300, 2018.

\bibitem{Rampazzo2018}
M.~Rampazzo, M.~Luvisotto, N.~Tomasone, I.~Fastelli, and M.~Schiavetti,
  ``Modelling and simulation of a {{Li}}-ion energy storage system: {{Case}}
  study from the island of {{Ventotene}} in the {{Tyrrhenian Sea}},''
  \emph{Journal of Energy Storage}, vol.~15, pp. 57--68, 2018.

\bibitem{Sakti2017}
A.~Sakti, K.~G. Gallagher, N.~Sepulveda, C.~Uckun, C.~Vergara, F.~J. {de
  Sisternes}, D.~W. Dees, and A.~Botterud, ``Enhanced representations of
  lithium-ion batteries in power systems models and their effect on the
  valuation of energy arbitrage applications,'' \emph{Journal of Power
  Sources}, vol. 342, pp. 279--291, 2017.

\bibitem{Pozo2014}
D.~Pozo, J.~Contreras, and E.~E. Sauma, ``Unit {{Commitment With Ideal}} and
  {{Generic Energy Storage Units}},'' \emph{IEEE Transactions on Power
  Systems}, vol.~29, no.~6, pp. 2974--2984, Nov. 2014.

\bibitem{Astaneh2018}
M.~Astaneh, R.~{Dufo-L{\'o}pez}, R.~Roshandel, F.~Golzar, and J.~L.
  {Bernal-Agust{\'i}n}, ``\BIBforeignlanguage{en}{A computationally efficient
  {{Li}}-ion electrochemical battery model for long-term analysis of
  stand-alone renewable energy systems},''
  \emph{\BIBforeignlanguage{en}{Journal of Energy Storage}}, vol.~17, pp.
  93--101, Jun. 2018.

\bibitem{Dong2018}
T.~Dong, P.~Peng, and F.~Jiang, ``\BIBforeignlanguage{en}{Numerical modeling
  and analysis of the thermal behavior of {{NCM}} lithium-ion batteries
  subjected to very high {{C}}-rate discharge/charge operations},''
  \emph{\BIBforeignlanguage{en}{International Journal of Heat and Mass
  Transfer}}, vol. 117, pp. 261--272, Feb. 2018.

\bibitem{DAmbrosio2010a}
C.~D'Ambrosio, A.~Lodi, and S.~Martello, ``\BIBforeignlanguage{en}{Piecewise
  linear approximation of functions of two variables in {{MILP}} models},''
  \emph{\BIBforeignlanguage{en}{Operations Research Letters}}, vol.~38, no.~1,
  pp. 39--46, Jan. 2010.

\bibitem{IEEE-RTS}
{Probability Methods Subcommittee}, ``{{IEEE}} reliability test system,''
  \emph{IEEE Transactions on Power Apparatus and Systems}, vol. PAS-98, no.~6,
  pp. 2047--2054, Nov. 1979.

\bibitem{GonzalezC_DataBatt}
A.~{Gonzalez-Castellanos}, D.~Pozo, and A.~Bischi,
  ``\BIBforeignlanguage{en}{Data for: {{A}} detailed {{Li}}-ion battery
  operation model},'' Oct. 2018, available at
  http://data.mendeley.com/datasets/36w7ts3r4t, version 1.

\bibitem{REE}
{Red El{\'e}ctrica de Espa{\~n}a}, ``Electricity demand tracking in real time,
  associated generation mix and {{CO2}} emissions for 06/20/2018,''
  https://demanda.ree.es/visiona/peninsula/demanda/total/2018-06-20.

\bibitem{Bezanson2017}
{Bezanson, J.}, {Edelman, A.}, {Karpinski, S.}, and {Shah, V.B.}, ``Julia: {{A
  Fresh Approach}} to {{Numerical Computing}},'' \emph{SIAM Review}, vol.~59,
  no.~1, pp. 65--98, Jan. 2017.

\bibitem{Lubin2015}
M.~Lubin and I.~Dunning, ``Computing in {{Operations Research Using Julia}},''
  \emph{INFORMS Journal on Computing}, vol.~27, no.~2, pp. 238--248, Apr. 2015.

\bibitem{gurobi}
{Gurobi Optimization Inc.}, ``Gurobi {{Optimization}}: The state-of-the-art
  mathematical programming solver,'' http://www.gurobi.com/.

\bibitem{IpOpt}
{A. W{\"a}chter} and {L. T. Biegler}, ``On the {{Implementation}} of a
  {{Primal}}-{{Dual Interior Point Filter Line Search Algorithm}} for
  {{Large}}-{{Scale Nonlinear Programming}},'' \emph{Mathematical Programming},
  vol. 106, no.~1, pp. 25--57, 2006.

\end{thebibliography}
\bibliographystyle{IEEEtran}


%

\begin{IEEEbiography}
[{\includegraphics[width=1in,height=1.25in,clip]{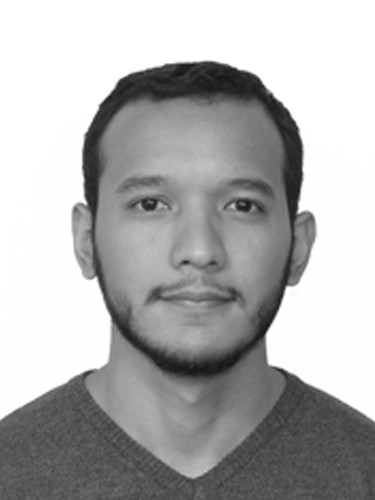}}]
{Alvaro Gonzalez--Castellanos} (S'11) received his B.Sc. in electrical engineering from the University of the North, Barranquilla, Colombia, in 2015, the M.Sc. in power engineering from the Skolkovo Institute of Science and Technology, Moscow, Russia, in 2017.
Currently, he's pursuing a Ph.D. with the Center for Energy Science and Technology at the Skolkovo Institute of Science and Technology.
His research interests include the convex characterization of flexibility measures in integrated energy infrastructures
\end{IEEEbiography}

\begin{IEEEbiography}
[{\includegraphics[width=1in,height=1.25in,clip]{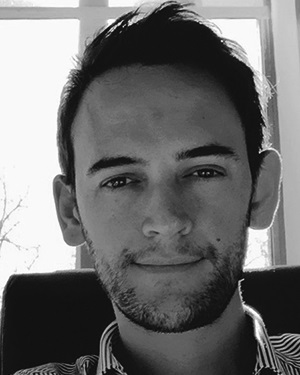}}]
{David Pozo} (S'06--–M'13--–SM'18) received his B.S. and Ph.D. degrees in electrical engineering from the University of Castilla-La Mancha, Ciudad Real, Spain, in 2006 and 2013, respectively. Since 2017, he is Assistant professor at the Skolkovo Institute of Science and Technology (Skoltech), Moscow, Russia. Prior to Skoltech Dr. Pozo worked as a postdoctoral fellow at the Pontifical Catholic University of Chile and the Pontifical Catholic University of Rio de Janeiro.
His research interest lies in the field of power systems and includes operations research, uncertainty, game theory, and electricity markets. He also focuses on problems of optimization and flexibility of modern power systems.
Since 2018, Prof. Pozo is leading the research group on Power Markets Analytics, Computer Science and Optimization (PACO).
\end{IEEEbiography}

\begin{IEEEbiography}[{\includegraphics[width=1in,height=1.25in,clip]{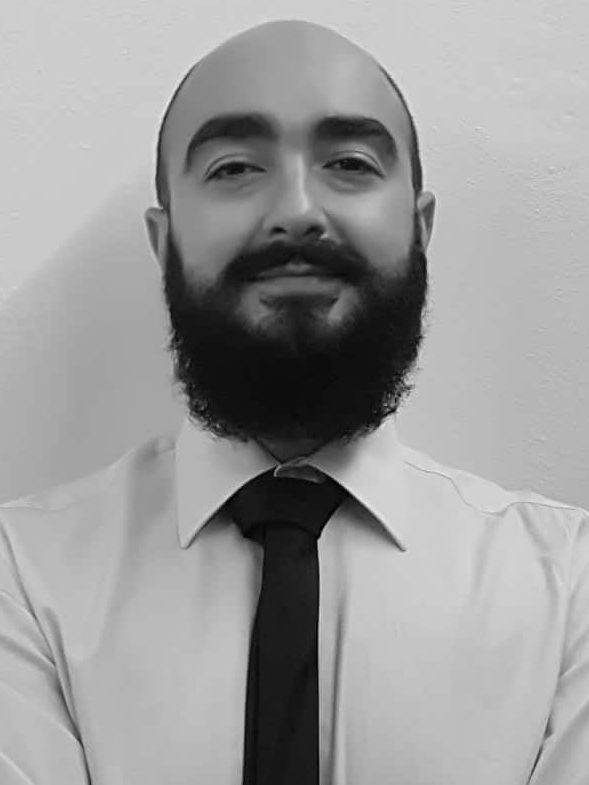}}]
{Aldo Bischi}, following M.Sc in Mechanical Engineering (2007) from University of Perugia (Italy) received his Ph.D. in Energy and Process Engineering (2012) from the Norwegian University of Science and Technology-NTNU (Norway). From 2012 until 2015 he held a post-doctoral position at Politecnico di Milano (Italy). Since 2016 he is Assistant Professor at Skolkovo Institute of Science and Technology (Russian Federation), part-time since 2018 when he started working as a postdoctoral research associate at University of Pisa (Italy).
Dr. Bischi scientific interest lies in the field of energy conversion systems, ranging from modeling and experimental activities to optimal scheduling and design, including all kinds of energy vectors such as gas, heat and electricity.
\end{IEEEbiography}





\vfill


\end{document}